\documentclass[prb,twocolumn,superscriptaddress,showpacs,amsmath,amssymb,floatfix]{revtex4-1}

\usepackage{graphicx}
\usepackage{latexsym}
\usepackage{amsmath}
\usepackage{amssymb}
\usepackage{amsfonts}
\usepackage{bm}
\usepackage{verbatim}
\usepackage{physics}

\newcommand{\Ione}{F}
\newcommand{\Itwo}{A}

\begin{document}
\title{Theory for the stationary polariton response in the presence of vibrations}
\date{\today}

\author{Kalle S. U. Kansanen}

 \affiliation{Department of
Physics and Nanoscience Center, University of Jyv\"askyl\"a, P.O.
Box 35 (YFL), FI-40014 University of Jyv\"askyl\"a, Finland}

\author{Aili Asikainen}

 \affiliation{Department of
Physics and Nanoscience Center, University of Jyv\"askyl\"a, P.O.
Box 35 (YFL), FI-40014 University of Jyv\"askyl\"a, Finland}

 \affiliation{Department of Computer Science, School of Science, Aalto University, FI-00076, Finland}

\author{J. Jussi Toppari}

 \affiliation{Department of
Physics and Nanoscience Center, University of Jyv\"askyl\"a, P.O.
Box 35 (YFL), FI-40014 University of Jyv\"askyl\"a, Finland}

\author{Gerrit Groenhof}

 \affiliation{Department of
Chemistry and Nanoscience Center, University of Jyv\"askyl\"a, P.O.
Box 35 (YFL), FI-40014 University of Jyv\"askyl\"a, Finland}

\author{Tero T.~Heikkil\"a}

 \affiliation{Department of
Physics and Nanoscience Center, University of Jyv\"askyl\"a, P.O.
Box 35 (YFL), FI-40014 University of Jyv\"askyl\"a, Finland}

\begin{abstract}
We construct a model describing the response of a hybrid system where
the electromagnetic field --- in particular, surface plasmon
polaritons --- couples strongly with electronic excitations
of atoms or molecules. Our approach is based on the input-output
theory of quantum optics, and in particular it takes into account the
thermal and quantum vibrations of the molecules. The latter is described within 
the $P(E)$ theory analogous to that used
in the theory of dynamical Coulomb blockade. As a result, we are able
to include the effect of the molecular Stokes shift on the strongly
coupled response of the system. Our model then accounts for the asymmetric emission from upper and lower polariton modes. It also allows for an accurate description of the partial decoherence of the light emission from the strongly coupled system. Our results can be readily used to connect the response of the hybrid modes to the emission and fluorescence properties of the individual molecules, and thus are relevant in understanding any utilization of such systems, like coherent light harvesting.
\end{abstract}

\maketitle

\section{Introduction}

Photonic structures, such as optical cavities or surface plasmon polaritons can modify electromagnetic vacuum field by confining the light to smaller volumes and restricting the number of available photonic modes. Any electronic excitation inside such modified vacuum can interact much stronger with the confined light mode. This interaction can become strong enough for the coupling energy to show up in absorption and emission spectra of such systems, suggesting formation of hybrid light-matter states, called polaritons. Common examples studied in this strong coupling limit are single atoms\cite{rempe1987observation}, excitons in semiconductors\cite{weisbuch1992observation}, and photoactive molecules~\cite{torma2014strong,yu2019strong}. 

More recently strong coupling of molecules with confined light modes has been in the focus of interest, because the hybridization between light and matter into polaritons not only delocalizes the excitation over
many molecules but also changes their potential energy surface, and thus provides a new way to control
chemistry\cite{hertzog2019strong}. 
Experiments on strongly coupled molecules have already shown 
(i) suppression of photo-oxidation of TBDC J-aggregates coupled to plasmonic nano-prisms\cite{munkhbat2018suppression}, and of photo-isomerization of Spiropyran inside an optical cavity \cite{hutchison2012modifying}; 
(ii) enhanced electronic conductivity in organic semiconductors\cite{orgiu2015conductivity};
(iii) inter-molecular excitation energy transfer over large distances inside optical cavities\cite{zhong2016non,zhong2017energy}; 
and (iv) enhanced decay of triplet states in Erythrosine B molecules\cite{stranius2018selective}. Since polaritons are like interacting dressed photons with mass they can undergo  Bose-Einstein condensation even at room temperature\cite{deng2010exciton}, which further enables very efficient and thresholdless polariton lasing\cite{deng2003polariton,christopoulos2007room}.

Strong coupling between a single molecule and electromagnetic field is very hard to achieve \cite{chikkaraddy2016single,wang2019turning}. 
The common way to circumvent this problem is to couple multiple molecules to the same photonic mode. Often these systems are still described within effective two-state models accounting only the two polaritonic states \cite{schwartz2013polariton}. However, such a description disregards the fact that the visible polariton modes are now superpositions of several molecular excitations and the photonic mode, and they are not the only eigenmodes of the system. The response of the whole system also depends on the presence of "dark modes", i.e. superpositions having no photonic component. These dark modes become relevant especially when dissipation processes within the molecules, such as those linked to vibrations, are included. In this case they can dramatically affect the predicted guiding of the chemistry, and even the validity of the whole concept. They have been taken into account in some multiscale simulations coupling the investigated molecules to thermal environments\cite{Groenhof2019JPCL}. 
However, those simulations often consider the transient response, whereas majority of experiments on light-matter coupling concern stationarily driven setups.

Here we construct a detailed description of the 
stationarily driven response of the strongly coupled system, taking into account the 
effect of inhomogeneous broadening of the
molecular response due to quantum and thermal vibrations of the molecules. 
We take the vibrations into account via the $P(E)$ theory
analogous to that used in Coulomb blockade~\cite{ingoldnazarov,terosbook}. 
This theory describes the probability of absorbing
(for $E>0$) or emitting ($E<0$) the energy $E$ to/from the
vibrations. For the specific models of harmonic vibrations, such a
$P(E)$ function can be calculated exactly. 
In general, we find how this $P(E)$ is related to
the absorption and emission spectra of individual molecules.
Therefore, an alternative approach is to deduce an effective $P(E)$ for the measured spectra of individual molecules.
The resulting fluorescence spectrum is similar
to that found via quantum many-body theory~\cite{mahan2013many}.
However, in these approaches only the transient response is considered which requires assumptions on the initial state of the system.
In our work, the focus is on the stationary response, in which case different conservation laws and their explicit breaking come into the focus.
What is more, we connect this fluorescence spectrum directly to the
absorption/emission spectrum of the strongly coupled system.
In a certain limit of parameters, the resulting inhomogeneous broadening of
the molecular absorption/emission then determines the linewidth of the
polariton modes.
In particular, our model explains the asymmetric emission spectra of upper and lower polaritons seen in many experiments \cite{baieva2017dynamics,bellessa2004strong,hakala2009vacuum,baieva2012strong,koponen2013absence,guebrou2012coherent}, as well as the varying polarization of that emission depending on the quantum coherence of the system as shown recently~\cite{baieva2017dynamics}.

\begin{figure}
	\includegraphics{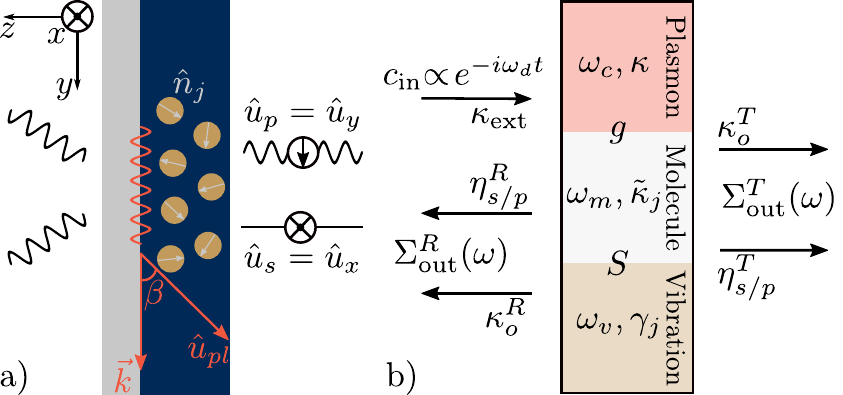}
    \caption{\label{fig:schematics} a) The measurement setup in which a surface plasmon polariton is excited on an interface where it can strongly couple to molecules. b)~Schematic with relevant parameters to the input-output formalism.}
\end{figure}

Besides the detailed description of the vibrations we include polarization of the confined light field and the positions of the molecules.
The position of a molecule with respect to the light field only amounts to a phase factor to the light-matter coupling but collectively it leads to experimentally observable effects.
Perhaps the most striking effect is the superradiance due to coherent emission described by Dicke in the 1950's\cite{dicke1954coherence}, but conversely it is seen in the usual experiments where many molecules are distributed over a region larger than the light wavelength.
On the other hand, the polarization of the light and the transition dipole moment of a molecule determine whether there is coupling at all: if these directions are perpendicular, the coupling vanishes.
This provides another way to control light-matter interaction which could be used in applications \cite{lodahl2017chiral}.
In this article we describe both the incoherent and coherent limits of polaritonics.

Although many aspects of our theory can be generalized to any confined light mode, like resonances of Fabry-P\'erot cavities, here we focus in particular on surface plasmon polaritons (below, plasmons) driven by an external light field. Plasmons are evanescent like electromagnetic modes propagating along metal-dielectric surface with a 2D momentum ${\vec k}$ along the surface. In general they have a non-linear dispersion $\omega({\vec k})$, and due to the evanescent nature their electromagnetic field is highly confined to the surface. Because of this confinement, the dipolar coupling to molecular excitations residing at the surface can be made strong \cite{bellessa2004strong, torma2014strong} leading to the observed avoided crossing between the two systems and thus offering possibility to control photochemical reactions. A typical way to launch plasmons is via Kretschmann configuration, i.e. coupling an external electromagnetic field to the surface modes via a prism \cite{kretschmann1968radiative}. In this setup the angle with which the light enters the prism determines a specific plasmon ${\vec k}$ vector. Hence, in this work we concentrate on a single plasmon mode with defined ${\vec k}$ and a generic frequency $\omega_c$. 

To be specific, we consider the plasmon--molecule system in the
strong-coupling regime. 
We describe the plasmon by a single bosonic mode~$c$ of frequency~$\omega_c$ and a given polarization~$\hat u_{pl}$ with respect to its wavevector~$\vec k$.
A concrete example of such a plasmon is the surface plasmon polariton traveling along an interface in the $xy$-plane in the $y$-direction with $\hat u_{pl} = \qty(0,\sin{\beta},\cos{\beta})$ as in Fig.~\ref{fig:schematics}a.
The plasmon interacts with $N$ identical molecules~\cite{scnote} which we approximate as two-level systems with transition frequency~$\omega_m$.
We denote the rising (lowering) operator of a molecule with $\sigma_j^\dagger$ ($\sigma_j$).
As in typical experiments,
we assume that the electric dipole moments of the molecules point in uniformly random directions~$\hat n_j$.
Following the standard approach of quantum optics \cite{walls2007quantum,gonzalez2013theory}, the Hamiltonian of the strong-coupled system is in the rotating wave approximation  ($\hbar = 1$)
\begin{equation}
	H_{s-c} = \omega_c c^\dagger c + \sum_{j=1}^N\qty( \omega_m \sigma_j^\dagger \sigma_j 
    +  g_j \sigma_j^\dagger c + g_j^* c^\dagger \sigma_j).
\end{equation}
The position~$\vec r_j$ of a molecule affects the coupling~$g_j$ in two ways: it contains a complex phase factor due to the phase of the plasmon, and the coupling strength depends on the distance to the interface.
If this distance is independent of the polarization the latter effect may be disregarded and the average value used.
Also, the coupling strength depends on the angle between the plasmon polarization and dipole moment of a molecule.
Thus, we write  $g_j = g e^{i \vec k \cdot \vec r_j}\qty(\hat n_j \cdot \hat u_{pl})$.

In addition to the strong-coupled system we include the vibrational modes of the molecules.
We assume a single vibration mode~$b_j$ per molecule with eigenfrequency~$\omega_v$ but the generalization to multiple modes is straightforward (Appendix~\ref{sec:multiplemodes}).
These vibrations and their interactions are described by
\begin{equation}
	H_v = \sum_{j=1}^N \omega_v b_j^\dagger b_j +  \sum_{j=1}^N \omega_v \sqrt{S} \sigma_j^\dagger \sigma_j \qty(b_j^\dagger + b_j).
	\label{eq:H:vm}
\end{equation}
The coupling between electronic and vibrational modes is quantified with a dimensionless parameter~$\sqrt{S}$, the Huang--Rhys factor~\cite{huang1950theory}, which is related to the Stokes shift measured in fluorescent emission.

We seek an approach to find the response of the strongly-coupled plasmon--molecule system in the presence of vibrations.
To this end, we employ the input-output formalism of quantum optics \cite{gardiner1985input,methodnote}.
We assume that there are separate bosonic baths for each molecule, vibration and the plasmon to which the coupling is linear in $\sigma_j, b_j$ and $c$, respectively.
In the Markov approximation these couplings are described by the dissipation rates  $\tilde \kappa_j, \gamma_j$, and $\kappa$ of the molecules, vibrations, and plasmon.
In the following, we suppose identical molecules and vibrations so that $\gamma_j = \gamma$ and $\tilde\kappa_j = \tilde\kappa$.
We neglect the thermal fluctuations of plasmons and molecules here as $\hbar \omega_m, \hbar\omega_c \gg k_B T$ even at room temperature.
We simplify the molecule--vibration Hamiltonian by introducing a new polaron operator $\sigma^S_j = e^{\sqrt{S}\qty(b_j^\dagger - b_j)}\sigma_j \equiv Q_j \sigma_j$.
Lastly, we assume a low driving power which corresponds to the single-excitation limit $\sigma^\dagger \sigma \approx 0$.
We find that when $\frac{\hbar \omega_v}{k_B T} > \frac{\gamma S}{2\kappa_m}$, where $\kappa_m = \tilde \kappa + \gamma S$ is the total effective damping rate of the individual molecules,  
the dynamics of the vibrational modes $b_j$ are approximately uncoupled from the plasmon--molecule system as shown in Appendix~\ref{sec:approximation}.
This allows us to use the Caldeira--Leggett model \cite{caldeira1984influence} for the vibrational dynamics. 
The plasmon and molecular equation are in this case
\begin{subequations}\label{eq:QLE}
\begin{align}
	\dot c &= - i \omega_c c - i \sum_j g_j^* \sigma_j^S Q_j^\dagger  
    - \frac{\kappa}{2} c - \sqrt{\kappa_{\rm ext}} c_{\rm in} \label{eq:QLE1}\\
	\dot \sigma_j^S &= - i \tilde{\omega}_m \sigma_j^S - i g_j Q_j c - \frac{\kappa_m}{2} \sigma_j^S - \sqrt{\kappa_m^{\rm ext}} Q_j\sigma_{{\rm in},j}, \label{eq:QLE2}
\end{align} 
\end{subequations}
where $\tilde\omega_m = \omega_m - S \omega_v$ is the renormalized molecular frequency while $\kappa_{\text{ext}}$ and $\kappa_m^{\rm ext}$ are the couplings to external driving fields.
For the plasmon--molecule system we assume that only the plasmon is driven so that $c_{\rm in} = \alpha e^{- i \omega_d t}$ and $\sigma_{{\rm in},j} = 0$.

We model a measurement on the plasmon--molecule system so that the incoming light $c_{\rm in}$ produces a reflected $\Sigma^R_{\rm{out}}$ and transmitted field $\Sigma^T_{\rm{out}}$.
These fields contain both the plasmon and the molecular emission but not the emission of phonons from the vibrations, because they are usually not measured. 
Phonon emission hence allows for a loss of energy in the process, so that the power in the output fields can be lower than the one in the input.
We also separately include coupling to s- and p-polarized light represented by $\hat u_p = \hat u_y$ and $\hat u_s = \hat u_x$ (Fig.~\ref{fig:schematics}a).
Since the propagating plasmon cannot emit s-polarized light to the direction perpendicular to the interface but the molecules have no directional preference, we consider s- and p-polarized output fields separately.
The output fields obey a general expression
\begin{equation}
	\Sigma_{\rm{out},s/p}^{T/R} = \qty(\delta_R^{T/R}c_{\rm in} + \sqrt{\kappa_o^{T/R}} c)\delta_p^{s/p} + \sum_j \eta_{j,s/p}^{T/R}\sigma_j.
    \label{eq:outputoperator}
\end{equation}
In this equation $\delta_R^{T} = 0$ and $\delta_R^{R} = 1$ meaning that only the reflected field interferes with the input field. 
The $\delta_p^{s/p}$ is defined similarly because the plasmon couples only to p-polarized modes.
The constants $\eta_{j,s/p}^{T/R}$ describe the coupling of the molecule electronic states to the environmental s- and p-polarized free space modes, and thus $\eta_{j,s/p}^{T/R} = \sqrt{\kappa_m^{T/R}}\qty(\hat n_j \cdot \hat u_{s/p})$.
These fields and couplings to the system are represented schematically in Fig.~\ref{fig:schematics}b.
The output spectral density is obtained from
\begin{equation}
	S^{T/R}_{s/p}(\omega;\omega_d) = \frac{1}{2 \pi}\!\int\!\dd{t} e^{i \omega t} \!\expval{\Sigma_{\rm{out},s/p}^{T/R\dagger}(0) \Sigma_{\rm{out},s/p}^{T/R}(t)}\!,
    \label{eq:psd}
\end{equation}
where $\omega$ is the frequency of the output field and $\omega_d$ the
driving frequency. 

We note that the Markov approximation leading to
Eqs.~ \eqref{eq:QLE} disregards the heating of the various baths of
the plasmons, molecules and the vibrations. These heating effects
can be disregarded when the heat conductance from those baths to
other degrees of freedom exceeds that due to the losses described by
$\kappa, \kappa_m$, and $\gamma$.

\section{$P(E)$ theory}\label{sec:PE}
 
The presence of the vibrations makes the input-output equations
\eqref{eq:QLE} non-linear as they contain products of different dynamical
fields. 
This non-linearity leads to an inelastic (fluorescent) response of the molecules to the light field, where the emitted light from the molecules takes place at lower frequencies than the absorption. This is often referred as the Stokes shift.
In order to take this non-linearity into account in the output spectra we introduce $P(E)$ theory similar to the one in dynamical Coulomb blockade~\cite{ingoldnazarov}. 
Recently, the same problem has been discussed in Ref.~\onlinecite{reitz2019langevin} using similar methods, but only in a specific limit of vibrations (see below). The identification of $P(E)$ allows for a more general approach enabling also resolving of polariton emission, which is lacking from Ref.~\onlinecite{reitz2019langevin}.
Let us first define $P(t) = \expval{Q_j^\dagger(t)Q_j(0)}$ and its Fourier transform
\begin{equation}
	P(E) = \frac{1}{2\pi} \int \dd{t} e^{i E t} P(t).
\end{equation} 
When the molecules are identical, $P(E)$ does not depend on the molecule index $j$; this assumption is easily lifted if needed.
The $P(E)$ function normalizes to unity and is real for stationary vibrations, i.e. $\expval{Q_j^\dagger(t + \tau)Q_j(\tau)} = \expval{Q_j^\dagger(t)Q_j(0)}$ for any time $\tau$. 
We can thus interpret the $P(E)$ function as a probability distribution of transforming energy $E$ to the vibrations ($E > 0$) or vice versa ($E < 0$).
This $P(E)$ function is characterized by four parameters: vibration eigenfrequency $\omega_v$, their linewidth $\gamma$, Huang--Rhys factor $S$, and temperature $T$ of their bath.
It constitutes a full description of the response of the vibrations.

We present a general derivation of the $P(E)$ function and a related $L$ function assuming Gaussian thermal fluctuations. 
Then, we derive $P(E)$ analytically in the limit in which $\gamma$ vanishes.
In this regime $P(E)$ is related to the absorption function defined by Huang and Rhys \cite{huang1950theory}. 
However, our analytic results for the response apply also in the case of general $\gamma$ and can be used for different models of vibrations.

\subsection{Derivation of $P(E)$}
We now derive the $P(E)$ function analytically in a similar manner as in the context of dynamical Coulomb blockade~\cite{ingoldnazarov}.
To establish notation, we omit the molecular index $j$ here and denote $x = b^\dagger + b$ and $p = i\qty(b^\dagger - b)$, the dimensionless position and momentum operator, respectively.
Then, we may write $Q^\dagger(t) = e^{i \sqrt{S}p(t)}$ in the correlator $P(t)$, which is the inverse Fourier transform of $P(E)$.
This correlator can be evaluated for thermal vibrations.
If the vibrations are described by a harmonic oscillator Hamiltonian, the
fluctuations are Gaussian and the weak version of the Wick's theorem
(see e.g. Ref.~\onlinecite{ingoldnazarov} and an example of a non-Gaussian $P(E)$ in Ref.~\onlinecite{heikkila04}) applies.
We identify $P(t)$ as the characteristic
function of fluctuations of the stochastic quantity $p(t)-p(0)$, where
everywhere in the calculations $p(t)$ should be ordered to the left of $p(0)$.
We assume the thermal vibrations to be stationary, and therefore the expectation value of $p(t)-p(0)$ vanishes (as $\expval{p(t)} = \expval{p(0)}$ for stationary vibrations).
Consequently for Gaussian fluctuations we can write the characteristic function in terms of the variance alone. 
In that case \cite{terosbook},
\begin{equation}
P(t)=e^{{\cal T} \langle (i\sqrt{S}[p(t)-p(0)])^2\rangle/2}=e^{S\langle
  [p(t)-p(0)]p(0)\rangle},
  \label{eq:Ptimecorr}
\end{equation}
where the latter equality uses the fact that $\langle p(t)^2\rangle =
\langle p(0)^2\rangle$. The operator ${\cal T}$ takes care of ordering
$p(t)$ before $p(0)$, but that operator is no longer needed in the
second equality because there $p(t)$ always precedes $p(0)$ in operator products.

Now, $p(t)$ can be obtained by solving
the quantum Langevin equations without rotating wave
approximation (also known as the Caldeira--Leggett model \cite{caldeira1984influence})
\begin{equation}
\begin{split}
\dot x(t) &= \omega_v p(t)\\
\dot p(t) &=-\omega_v x(t) -\gamma p(t) + \xi(t),
\end{split}
\label{eq:langevineq}
\end{equation}
where $\gamma$ is the linewidth of vibrations, and $\xi$ is a Langevin
force describing the thermal fluctuations. It has the correlator
\begin{equation}
\langle \xi(t) \xi(t') \rangle = \int \dd{\omega}
\exp[-i\omega(t-t')] S_\xi(\omega),
\label{eq:noisecorr}
\end{equation}
where the noise correlator is given by
\begin{equation}
S_\xi(\omega)=  \frac{\gamma\omega}{\pi\omega_v}
\left[\coth\left(\frac{\omega}{2k_B T}\right)+1\right]
\label{eq:noiseS}
\end{equation}
for thermal noise \cite{giovannetti2001phase}.

The Langevin equations \eqref{eq:langevineq} can be solved
via Fourier transform. The result is
\begin{equation}
\begin{pmatrix}x(\omega)\\p(\omega)\end{pmatrix} =
\frac{1}{\omega^2-\omega_v^2+i\omega \gamma} \begin{pmatrix}
  -\omega_v\\ i\omega\end{pmatrix} \xi(\omega).
  \label{eq:result:xp}
\end{equation}
After some Fourier analysis with the help of Eqs.~\eqref{eq:noisecorr},~\eqref{eq:noiseS} and \eqref{eq:result:xp} we find 
$P(t)=e^{J(t)-J(0)}$  according to Eq.~\eqref{eq:Ptimecorr} with
\begin{widetext}
\begin{equation}
J(t) = S\expval{p(t)p(0)} =\frac{S \gamma}{\pi \omega_v} \int \dd{\omega} e^{-i\omega t}
  \frac{\omega^3}{(\omega^2-\omega_v^2)^2+\omega^2 \gamma^2}
    \left[\coth\left(\frac{\omega}{2k_B T}\right)+1\right].
\label{eq:Joft}
\end{equation}
\end{widetext}
The resulting $P(E)$ is thus governed by three dimensionless
parameters: the Huang--Rhys factor $S$, the quality factor of
vibrations $\omega_v/\gamma$, and the relative temperature $k_B T/\omega_v$.
Note that in the Caldeira--Leggett model $J(t)$ is related to the vibrational spectral density $J_v(t) = S \omega_v^2 \expval{x(t)x(0)}$ via their respective Fourier transforms by $J_v(\omega) = J(\omega)/\omega^2$. 

A simpler expression for $J(t)$ is obtained if instead of the Caldeira--Leggett model one uses the usual quantum optical equation $\dot b = -(i\omega_v + \frac{\gamma}{2})b + \sqrt{\gamma}b_{\rm in}$ as in \cite{reitz2019langevin} for example. 
Then for white noise $\expval{b_{\rm in}(t)b_{\rm in}^\dagger(t')} = \qty(n_{\rm th}+1)\delta(t - t')$ we find
\begin{equation}
   J(t) = S(n_{\rm th} + 1)e^{-i\omega_v t - \frac{\gamma}{2} \abs{t}} + S n_{\rm th} e^{i\omega_v t - \frac{\gamma}{2} \abs{t}}.
   \label{eq:J:whitenoise}
\end{equation}
Here $n_{\rm th} = \qty(e^{\omega_v/(k_B T)} - 1)^{-1}$ is the Bose factor, i.e. the mean number of thermal phonons at the vibrational frequency $\omega_v$.
We arrive at the same solution from the Caldeira--Leggett model by using the method of residues to calculate the integral \eqref{eq:Joft} and then approximating $\gamma \ll \omega_v$.
This is hence the limit where Ref.~\onlinecite{reitz2019langevin} is valid. 
However, typical multiscale quantum chemistry calculations assume the opposite limit of a large $\gamma \gtrsim \tilde\kappa$, where molecular vibrations decay before the photon excitation.

Lastly, there are two general properties of the $P(E)$ function worth noting.
First, since $P(t)$ may be regarded as a characteristic function of the probability distribution $P(E)$, the raw moments of the energy can be expressed as
\begin{equation}
    \mathbb{E}(E^n) = i^n \eval{\dv[n]{P(t)}{t}}_{t=0} 
    = i^n \eval{\dv[n]{e^{J(t)}}{t}}_{t=0}.
    \label{eq:genEmom}
\end{equation}
With the help of this formula the mean and variance of $P(E)$ can be found.
Second, the Kubo--Martin--Schwinger (KMS) relation
for thermal fluctuations at temperature $T$ leads to the detailed balance condition (or
emission-absorption asymmetry) for $P(E)$
\begin{equation}
    P(-E)=\exp\left(-\frac{E}{k_B T}\right) P(E).
\end{equation}
This asymmetry in $P(E)$ is relevant for the anti-Stokes part of the spectrum.
Some approximations, such as the white noise approximation, break this balance condition.

\subsection{$L$ function}
Another quantity we encounter that is relevant for the emission spectrum of a molecule is the Fourier transform of the four-point correlator
\begin{equation}
    L(t_1,t_2,t_3) = \expval{Q^\dagger(t_1)Q(t_2)Q^\dagger(0)Q(t_3)}.
    \label{eq:L:start}
\end{equation}
This function is clearly related to $P(t)$ as for certain time arguments it coincides with the definition of $P(t)$, e.g. $L(t,0,0) = P(t)$.
Using the same assumptions as in the derivation of $P(E)$ we may write
\begin{equation}
    L(t_1,t_2,t_3) = e^{-S{\cal T} \langle [p(t_1)-p(t_2)+p(0) -p(t_3)]^2\rangle/2},
    \label{eq:L:2}
\end{equation}
where $\cal T$ orders operator products so that they are in the same order as in Eq.~\eqref{eq:L:start}.
Now, since the vibrations are stationary, we may write $L$ in terms of $P(t)$'s
\begin{equation}
    L(t_1,t_2,t_3) = \frac{P(t_1 - t_2)P(t_1 - t_3)P(t_2)P(-t_3)}{P(t_1)P(t_2-t_3)}.
    \label{eq:L:result}
\end{equation}
Even if we can fully calculate $J(t)$, the Fourier transform of $L$ is not straightforward to evaluate numerically in the general case.

\subsection{$\gamma = 0$ limit}
Next, we consider the limit in which the dissipation rate of vibrations vanishes and derive expressions for both $P(E)$ and $L$.
We note that the definition of $J(t)$, Eq.~\eqref{eq:Joft}, contains a nascent delta function
\begin{equation}
	\tilde{P}(\omega) = \frac{1}{\pi} \frac{\gamma \omega^2}{(\omega_v^2-\omega^2)^2+\omega^2 \gamma^2}
	\label{eq:twopeakLorentz}
\end{equation}
which in the limit $\gamma \rightarrow 0$ reduces to $\tilde{P}(\omega) = \frac{1}{2}\qty(\delta(\omega - \omega_v) + \delta(\omega + \omega_v))$.
Therefore
\begin{equation}
    J(t) =  S\qty(n_{\rm th} + 1)e^{-i \omega_v t} + S n_{\rm th} e^{i \omega_v t}.
    \label{eq:Jtgamma0}
\end{equation}
Note that this coincides with the limit $\gamma \rightarrow 0$ in the white noise model Eq.~\eqref{eq:J:whitenoise}.
The corresponding characteristic function $P(t)$ is known in probability theory to be that of the \emph{Skellam distribution} \cite{skellam1946frequency}.
It is a distribution that describes the difference of two independent Poisson processes.
In our case, these processes are the emission and absorption of phonons.
$P(E)$ then describes the total number of phonons transferred from/to vibrations to/from their environment. 
The resulting $P(E)$ function is
\begin{subequations}
\label{eq:PEgamma0}
\begin{align}
	P(E) &= \sum_{k=-\infty}^\infty p_k(S) \delta(E - k \omega_v) \\
	p_k(S) &= e^{-S(2 n_{\rm th} + 1)} \qty(1 + \frac{1}{n_{\rm th}})^\frac{k}{2} I_k\qty(2S\sqrt{n_{\rm th}(n_{\rm th}+1)}),
\end{align}
\end{subequations}
where $I_k(x)$ is the modified Bessel function of the first kind.
In the zero-temperature limit $p_k(S) = e^{-S}\frac{S^k}{k!}$ for $k\geq 0$ and $p_k(S) = 0$ for $k < 0$, i.e. the probability to emit phonons becomes Poissonian and the absorption probability vanishes.

We find the average and variance of the $\gamma=0$ distribution by using Eqs.~\eqref{eq:genEmom} and~\eqref{eq:Jtgamma0}
\begin{equation}
    \mathbb{E}(E) = S \omega_v 
    \qq{and} 
    \text{var}(E) = (2 n_{\rm th} + 1)S \omega_v^2.
\end{equation}
The variance depends on the temperature so that for high temperatures $k_B T \gg \omega_v$ the variance is directly proportional to the temperature; $\text{var}(E) \approx 2 S \omega_v k_B T$.
It should be noted that both the variance and the average are proportional to $S$ which also holds for a Poissonian quantity.
The physical picture is that the mean of $E$ describes the Stokes shift in the molecules whereas the variance (or standard deviation) is connected with the inhomogeneous broadening of the molecular linewidth due to vibrations.

Finally, we derive $L$ in the $\gamma \rightarrow 0$ limit using Eq.~\eqref{eq:L:result}.
It is necessary to simplify $1/P(t)$ in order to find the Fourier transform of $L$. 
Since $P(t) = e^{J(t) - J(0)}$ and $J(t) \propto S$ we may find $1/P(t)$ by changing $S \rightarrow -S$ in Eq.~\eqref{eq:PEgamma0}.
Using the parity of the modified Bessel function of the first kind $I_k(-x) = (-1)^k I_k(x)$ we can express the inverse as
\begin{widetext}
\begin{align}
    1/P(t) &= \sum_{k=-\infty}^\infty p_k(-S) e^{- i k \omega_v t} 
    = \sum_{k=-\infty}^\infty (-1)^k  \exp[2 S(2 n_{\rm th} + 1)] p_k(S) e^{- i k \omega_v t}.
    \label{eq:inversePtgamma0}
\end{align}
Below, we omit the $S$-dependence and denote $p_k(S) = p_k$.

The Fourier transform of $L$ is straightforward with the help of Eq. \eqref{eq:inversePtgamma0}.
We obtain

\begin{align}
    L(\omega_1,\omega_2,\omega_3) &= \frac{1}{(2 \pi)^3} \int \dd{t_1}\dd{t_2}\dd{t_3}L(t_1,t_2,t_3) 
    e^{i \omega_1 t_1 + i \omega_2 t_2 + i \omega_3 t_3}  \notag\\
   &= \sum_{k_1,k_2,k_3,k_4,k_5,k_6} (-1)^{k_1+k_2} p_{k_1}p_{k_2}p_{k_3}p_{k_4}p_{k_5}p_{k_6}
    e^{4S(2 n_{\rm th} + 1)}
    \delta(\omega_1-\qty[k_1+k_3+k_4]\omega_v) \notag\\
    &\phantom{\sum_{k_1,k_2,k_3,k_4,k_5,k_6}}\times\delta(\omega_2 - \qty[k_2 - k_3 + k_5]\omega_v)\delta(\omega_3 + \qty[k_2 + k_4 + k_6]\omega_v).
    \label{eq:L:gamma0}
\end{align}

This result can be used to obtain the fluorescence spectrum of a molecule.
The expression is slightly cumbersome to use because the six sums obtain values from $-\infty$ (or from $0$ when $T=0$) to $\infty$.
This problem is alleviated by the rapid decrease of $p_k$ as a function of $k$. Consequently, Eq.~\eqref{eq:L:gamma0} is straigthforward to compute numerically.
\end{widetext}

\begin{figure}
	\includegraphics{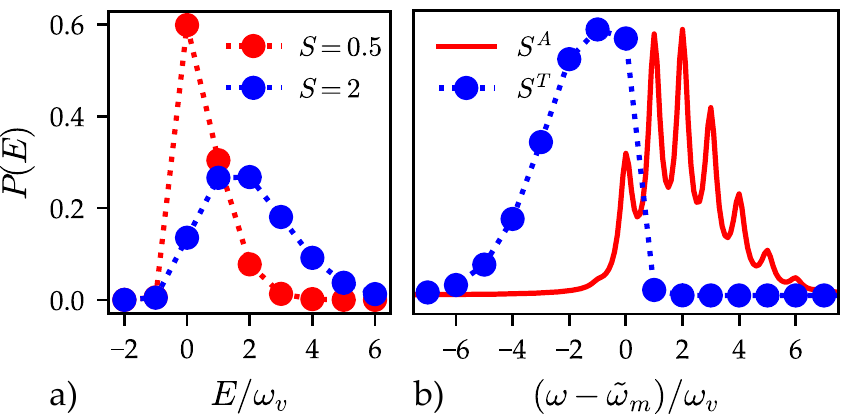}
    \caption{\label{fig:molspectra} a) $P(E)$ functions for $\gamma = 0$ showing the weights of $\delta$-function peaks when $\frac{\kappa_m}{\omega_v} = 0.5$ and $\frac{k_B T}{\omega_v} = 0.5$. b) The normalized emission and absorption with $S=2$. Emission is evaluated with the driving frequency $\omega_d = \tilde{\omega}_m$. Here, we choose $\kappa_m^{T/R} = \kappa_m^{\rm ext} = \frac{\kappa_m}{10}$.}
\end{figure}

\section{Stokes shift}\label{sec:stokesshift}
Before solving the full plasmon--molecule problem we illustrate how the $P(E)$ theory is used to model a measurement of the Stokes shift in a molecule--vibration system.
This is achieved by removing the plasmon term from Eq.~\eqref{eq:QLE2} and driving the molecules, i.e., adding the term
$\sigma_{{\rm in},j} = \frac{\alpha e^{i\theta_j}}{\sqrt{N}}\delta(\omega-\omega_d)$
where $\theta_j$ represents the phase of the driving field for molecule $j$.
The driving is scaled so that the total input power spectral density is given by~$I_{\rm in}=\abs{\alpha}^2\delta(\omega - \omega_d)$.
Then, we solve Eq.~\eqref{eq:QLE2} with Fourier transform and convolution theorem.
The spectra~$S^{T/R}$ are found from Eq.~\eqref{eq:psd} when the output fields are changed to 
$\Sigma^{T/R}_{\rm out} = \sum_j(\sqrt{\kappa_m^{T/R}}\sigma_j +\delta^{T/R}_R\sigma_{{\rm in},j})$.
Here, the 'reflected' field should not be understood literally but rather as the field that contains the driving field. 
The 'transmitted' field is fully from the molecular fluorescence.
Since $\sigma_j = Q_j^\dagger \sigma_j^S$ and the solution $\sigma_j^S$ of Eq.~\eqref{eq:QLE2} depends on $Q_j$ we encounter a four-point correlator $\expval{Q^\dagger_j(\omega_1) Q_j(\omega_2) Q_k^\dagger(\omega_3) Q_k(\omega_4)}$ in the calculation of $S^{T/R}$. 
Here, $Q^\dagger_j(\omega)$ refers to the Fourier transform of $Q_j^\dagger(t)$.
Assuming that the vibration modes are independent and identical in different molecules, the correlator factorizes into two-point correlators when $j \neq k$.
These resulting two- and four-point correlators are related to $P(E)$ by
\begin{subequations}
\begin{align}
	&\expval{Q_j^\dagger(\omega_1) Q_j(\omega_2)} = P(\omega_1) \delta(\omega_1 + \omega_2) \label{eq:2pointcorr}\\
    &\expval{Q_j^\dagger(\omega_1) Q_j(\omega_2)  Q_j^\dagger(\omega_3) Q_j(\omega_4)} \notag\\ 
    &\qquad = L(\omega_1,\omega_2,\omega_4) \delta(\omega_1 +\omega_2 + \omega_3 + \omega_4), \label{eq:4pointcorr}
\end{align}
\end{subequations}
where $L(\omega_1,\omega_2,\omega_4)$ is the Fourier transform of Eq.~\eqref{eq:L:result} in the general case.

When discussing the response of molecules to driving it is useful to introduce a frequency $\Delta = \omega_d - \tilde{\omega}_m$  which is the detuning between the driving and renormalized molecular frequency.
Without vibrations the molecular response is characterized by $\chi(\Delta) = (i\Delta - \frac{\kappa_m}{2})^{-1}$ which describes Lorentzian absorption and emission spectra.
However, in the presence of vibrations, the information about the spectral properties is contained in 
\begin{equation}
	\Itwo(\Delta) = \int \dd{E} P(E)\chi(\Delta - E)
    \label{eq:I2}
\end{equation}
and
\begin{equation}
	\Ione = \int \!\dd{\omega_1}\!\dd{\omega_2} L(\omega_1,\omega_d-\omega-\omega_1,\omega_2) \chi(\omega_1 - \Delta)  \chi(\omega_2 +\Delta).
    \label{eq:I1}
\end{equation}
These functions are associated with absorption and fluorescence of molecules, respectively, and play an important role in the plasmon--molecule problem.

\subsection{Incoherent limit}
Let us assume that molecules are randomly arranged, so that the phase $\theta_j$ is random. Averaging over them, 
the resulting spectra are
\begin{align}
    &\frac{S^{T/R}(\omega;\omega_d)}{\abs{\alpha}^2} = 
    \kappa_m^{T/R}\kappa_m^{\rm ext} \Ione(\Delta; \omega - \omega_d) 
    \label{eq:moleculespectra}\\
    &\quad + \delta_R^{T/R}\qty[1 + 2\sqrt{\kappa_m^{T/R} \kappa_m^{\rm ext}}\Re{\Itwo(\Delta)}]\delta(\omega-\omega_d). 
    \notag
\end{align}
The emission spectrum $S^T$ is determined by $\Ione$ which describes inelastic scattering (output field frequency $\omega$ different from driving frequency $\omega_d$). 
In the 'reflected' field $S^R$ we find also the input power spectral density $I_{\rm in}$ and a term proportional to $A$ representing absorption.

Both $\Ione$ and $\Itwo$ are straightforward to determine from Eqs. \eqref{eq:PEgamma0} and \eqref{eq:L:gamma0}, i.e., when the vibrational linewidth $\gamma \rightarrow 0$.
Then, $\Ione$ is also delta-peaked at frequencies $\omega - \omega_d = m \omega_v$ with an integer~$m$.
The absorption spectrum is obtained from power conservation $S^A(\omega_d) = I_{\rm in} - S^T - S^R$ evaluated at the driving frequency $\omega = \omega_d$ and it is mostly determined by $\Itwo$.
In Fig.~\ref{fig:molspectra}b we have plotted the emission spectrum $S^T$ along with the absorption spectrum $S^A$.
The absorption maximum is at the bare molecular frequency $\omega_m$ while the emission maximum is at approximately $\omega_m - 2S\omega_v$.
The difference is the Stokes shift.
The spectra correspond to the results describing the transient response obtained with Green functions~\cite{mahan2013many}. 
However, in our stationary model the absorption is not a mirror image of the emission because the emission may happen also from the excited vibrational states.

\subsection{Coherent limit}
Besides the experimentally more typical incoherent situation we look at the coherent limit.
Then the phase $e^{i \theta_j}$ is fixed.
This happens for instance when the distance between the molecules is much smaller than the wavelength of the driving field
or the molecules are in a suitably chosen lattice.
We renormalize the input in this case to be $\sigma_{{\rm in},j} = \frac{\alpha}{N} e^{-i \omega_d t}$ so that again the total input power is distributed evenly and is independent of the number of molecules $N$.
Then, the calculation can be repeated to give
\begin{align}
    &\frac{S^{T/R}(\omega;\omega_d)}{\abs{\alpha}^2} = 
    \delta^{T/R}_R\abs{1 + \sqrt{\kappa_m^{T/R}\kappa_m^{\rm ext}}\Itwo(\Delta)}^2\delta(\omega - \omega_d) \notag\\ 
    & + \kappa_m^{T/R}\kappa_m^{\rm ext} \qty[\frac{\Ione}{N} + \qty(\delta^{T/R}_T - \frac{1}{N})\abs{\Itwo(\Delta)}^2\delta(\omega - \omega_d)].
\end{align}
Interestingly, we obtain an explicit dependence on the number $N$ of molecules for two terms.
One of those terms is the inelastic emission term $\Ione$ which means that for large $N$ the spectra are mostly elastic.
However, if the vibrations are absent, i.e. $S=0$ which leads to $P(E) = \delta(E)$, $\Ione = \abs{\Itwo(\Delta)}^2\delta(\omega - \omega_d)$ and the $1/N$ dependent terms cancel. 
Therefore, this coherent effect is not related to sub- or superradiance of molecules described by Dicke \cite{dicke1954coherence}. Rather, it is related to vibrations and their enhanced non-radiative emission which shows up as a diminishing fluorescence as the number of molecules increases.

\section{Plasmon--molecule system}
\begin{figure}
	\includegraphics{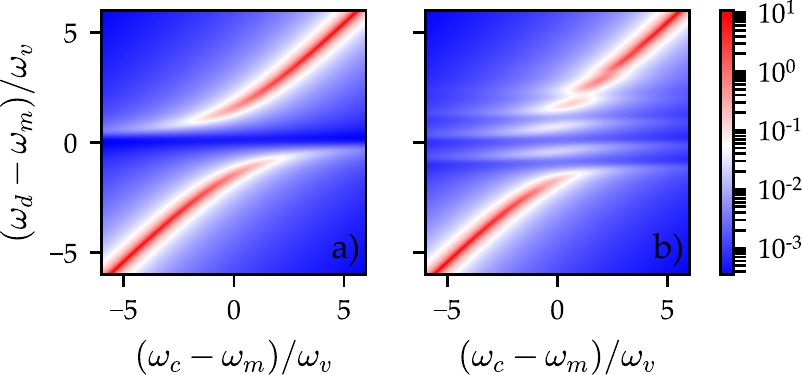}
    \caption{\label{fig:rfs} Response function $\abs{r(\omega_d)}^2/\omega_v$ of Eq.~\eqref{eq:response} for a single molecule for a) $S=0$ and b) $S=1$. The other parameters are $g/\omega_v = 1.5$, $\frac{k_B T}{\omega_v} = 0.5$, $\kappa/\omega_v = 0.1$, $\kappa_{\rm ext} = \kappa/2$, and $\kappa_m/\omega_v=0.5$. The response function also determines the direct plasmon emission spectrum.}
\end{figure}

With the tools developed in Sec.~\ref{sec:PE} and~\ref{sec:stokesshift} we can return to the problem of strongly coupled plasmon--molecule system and find the polarized spectra of the system using Eqs.~\eqref{eq:QLE} (with only the plasmon being driven, i.e. $\sigma_{{\rm in},j} = 0$ in this case).
We integrate Eq.~\eqref{eq:QLE2} from an initial time $t_i \rightarrow -\infty$ to $t_f=t$ and neglect the initial condition $\sigma_j^S(t_i)$ which has no role in a stationary situation.
We substitute this into Eq.~\eqref{eq:QLE1} which leads to
\begin{align}
	\dot c =& -\qty(i \omega_c + \frac{\kappa}{2})c - \sqrt{\kappa_{\rm ext}}c_{\rm in} \\
    &- \sum_j \abs{g_j}^2 \int^t_{-\infty} \dd{t'} e^{\qty(i \tilde{\omega}_m + \frac{\kappa_m}{2})\qty(t'-t)} Q_j^\dagger(t) Q_j(t') c(t').\notag
\end{align}
At this point we average the equation over the fluctuating vibrations and use a mean-field approximation. 
This leads to the $P(E)$ function since $\expval{Q_j^\dagger(t) Q_j(t')} c(t') = P(t-t')c(t')$.
Consequently, the elastic response of the plasmon is given by $c(t) = \alpha r(\omega_d) e^{-i \omega_d t}$ where 
\begin{equation}
	\frac{r(\omega_d)}{\sqrt{\kappa_{\rm ext}}} = \qty[i(\omega_d - \omega_c) - \frac{\kappa}{2} + \sum_j \abs{g_j}^2 \Itwo(\Delta)]^{-1}.
    \label{eq:response}
\end{equation}
Vibrations provide a channel of relaxation broadening the response which is associated with the real part of $\Itwo$.
The imaginary part contains information about the frequencies of the polariton modes.
When the vibrations are absent, $S=0$ and $P(E)=\delta(E)$, the usual strong-coupling response is obtained as $\Itwo \rightarrow \chi$ with Rabi splitting proportional to $\sqrt{\sum_j \abs{g_j}^2}$ at $\omega_c = \omega_m$. 
When the vibrations are present, especially the upper polariton branch is perturbed as Fig.~\ref{fig:rfs} shows.

Finally, $\sigma_j^S$ can be solved from Eq.~\eqref{eq:QLE2}  in terms of $Q_j$ by Fourier transformation using the convolution theorem and $c(\omega) = \alpha r(\omega_d) \delta(\omega - \omega_d)$.
Then we have all we need to evaluate the output spectra with Eqs.~\eqref{eq:outputoperator} and~\eqref{eq:psd}.

\subsection{Incoherent polaritonic response}
Let us consider a large number $N$ of identical molecules with random dipole moment directions $\hat n_j$.
In this case we can replace the sums over the molecule index with an integral over a surface of a sphere $\sum_j \rightarrow \frac{N}{4\pi}\int \dd{\Omega}$.
Then, because $g_j = g e^{i \vec k \cdot \vec r_j}\qty(\hat n_j \cdot \hat u_{pl})$ where $\hat u_{pl} = \qty(0,\sin{\beta},\cos{\beta})$ is the plasmon polarization vector, 
the square of the Rabi splitting in Eq.~\eqref{eq:response} is 
$\sum_j \abs{g_j}^2 = N g^2/3 \equiv g_N^2$.
We assume that the positions of the $N$ molecules are random over a region large compared to the wavelength of the plasmon so that we may replace $e^{i \vec k \cdot (\vec r_j - \vec r_k)} \rightarrow \delta_{jk}$ for an ensemble average.
Using these assumptions the polarization dependence shows up in the spectra as the coefficients
\begin{equation}
    C_{s/p} = \sum_{j,k} g_j \eta_{j,s/p}^{T/R} \qty(g_k \eta_{k,s/p}^{T/R})^{\!*} =
        \begin{cases} 
             \frac{\kappa_m^{T/R}g_N^2}{5}  \\
             \frac{\kappa_m^{T/R}g_N^2}{5} [2 - \cos{2\beta}], 
        \end{cases}
\end{equation}
where the upper/lower line is for s/p.
Above, only the terms where $j=k$ contribute in the sum which results in four-point correlators as in Eq.~\eqref{eq:4pointcorr}.

The s- and p-polarized emission spectra are
\begin{subequations}\label{eq:scemission}
\begin{align}
	S^T_s(\omega;\omega_d) &= \abs{\alpha r(\omega_d)}^2  C_s \Ione
	\label{eq:semission} \\
	S^T_p(\omega;\omega_d) &=  \abs{\alpha r(\omega_d)}^2 \qty[\kappa_o^T \delta(\omega - \omega_d)
    + C_p \Ione].
    \label{eq:pemission}
\end{align}
\end{subequations}
The main difference between the s- and p-polarized spectra is because the plasmon emits only p-polarized light.
The molecular fluorescence is also slightly enhanced in this polarization for $\beta \neq 0$.

Both the s- and p-polarized emission spectra are now represented with $\Ione$ and $\Itwo$ found in molecular fluorescence Eq.~\eqref{eq:moleculespectra}. 
Therefore, the emission of strongly coupled plasmon--molecule system is related to the properties of the plasmon and the molecules separately with a few parameters describing the plasmon--molecule coupling strength and their intrinsic decay rates.
Note also that these results hold for any $P(E)$, i.e., these results are independent of the vibrational model.

Since only the terms with $\Ione$ contribute to the inelastic emission, for a given driving frequency $\omega_d$ the ratio of p- and s-polarized emission at $\omega \neq \omega_d$ is  $S^T_p/S^T_s = 2 - \cos(2\beta)$.
For example, for an interface of vacuum and silver in the Drude model~\cite{yang2015optical,gonzalez2013theory} $\beta \approx \frac{\pi}{12}$ and $S^T_p/S^T_s \approx 1.13$ at $\omega_c = 2\,\rm{eV}$.
This ratio is otherwise independent of the system.

\begin{figure}
	\includegraphics{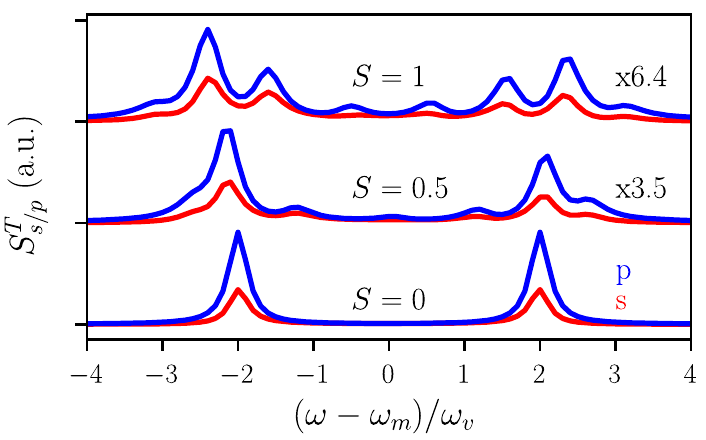}
    \caption{\label{fig:elemission} Elastic emission spectra for s- and p-polarization (red and blue curves, respectively) from a plasmon--molecule system with $\omega_c = \omega_m$. The different curves are offset and scaled for clarity. 
    The parameters are the same as in Fig.~\ref{fig:rfs} except for $\frac{g_N}{\omega_v} = 2$, $\beta = \frac{\pi}{12}$, $\kappa_o^T = \kappa/2$, $\kappa_m^T = \kappa_m/3$ and $\frac{k_B T}{\omega_v}=1$.}
\end{figure}

In Fig.~\ref{fig:elemission} we have plotted the elastic emission spectra from Eqs.~\eqref{eq:scemission} using the $P(E)$ function~\eqref{eq:PEgamma0}.
The s- and p-polarized emission are similar for small $S$ but for larger values the competition between the plasmon and molecule emission becomes more noticeable. 
The ratio between elastically emitted p- and s-polarized power is controlled by the ratio between $\kappa_o^T$ and $C_{s/p}$ and the detuning $\omega_c - \omega_m$.

We find that the upper and lower polariton modes emit asymmetrically as in other approaches~\cite{neuman2018origin,delpino2018tensor,herrera2017theory, Herrera2017PRL,Herrera2017PRA,zeb2017exact,reitz2019langevin,Michetti2008PRB} and experiments~\cite{baieva2017dynamics,bellessa2004strong,hakala2009vacuum,baieva2012strong,Chovan2008PRB}.
This is caused by the asymmetry of the effective dissipation rate $\Gamma=\kappa/2 - g^2_N \Re{\Itwo(\Delta)}$ which is related to the molecule's absorption spectrum.
The number of molecules affects the dissipation rate only via the size of the Rabi splitting.
Analytical insight can be obtained when $k_B T \ll \hbar \omega_v$ and $S \ll 1$. 
Then, we may consider only single-phonon processes.
When $g_N > \omega_v$, we find the polariton frequencies from the response function~\eqref{eq:response} for zero detuning $\omega_c = \omega_m$ to be approximately
\begin{equation}
    \omega_\pm = \omega_m \pm g_N + \frac{S\omega_v}{2}\frac{1}{\pm g_N/\omega_v - 1}.
\end{equation}
Thus, the vibrations affect both the position of the polariton peaks as well as the size of the Rabi splitting.
At these frequencies the dissipation rate in the first order of $\kappa_m$ is given by
\begin{equation}
    \Gamma_\pm=\frac{\kappa}{2} + \frac{\kappa_m}{2}\qty[1 +  \frac{S\omega_v^2}{(g_N\mp\omega_v)^2} + \frac{S\omega_v^2}{g_N(g_N\mp\omega_v)}].
\end{equation}
Due to vibrations the dissipation rate of the upper polariton ($\Gamma_+$) is larger than the dissipation rate of the lower polariton ($\Gamma_-$) which suppresses the upper polariton emission compared to the lower polariton.

\begin{figure}
	\includegraphics{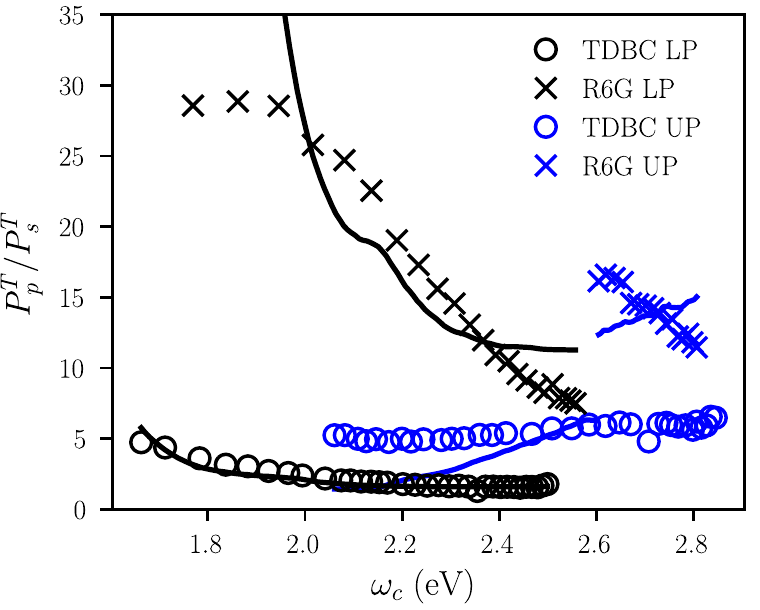}
    \caption{\label{fig:comparison} Comparison of experimental polarization ratio data and corresponding theoretical fits for two different molecules. For TDBC $\omega_m \approx 2.10\,\mathrm{eV}$ and $2S\omega_v \approx 5\,\mathrm{meV}$ and for R6G $\omega_m \approx 2.27\,\mathrm{eV}$ and $2S\omega_v \approx 97\,\mathrm{meV}$. We estimate the plasmon linewidth to be $\kappa = 250\,\mathrm{meV}$. The Rabi splitting for TDBC is $167\,\mathrm{meV}$ and for R6G $337\,\mathrm{meV}$. Our data on fluorescence of TDBC is limited below $2.6\,\mathrm{eV}$ so we cannot produce an estimate for the polarization ratio above $2.6\,\mathrm{eV}$. In the legend LP (UP) refers to lower (upper) polariton.} 
\end{figure}

An alternative method of using the equations for polarized emission spectra~\eqref{eq:scemission} is to use experimental molecular absorption and fluorescence data.
Then, to a good accuracy the lineshape of absorption is related to $\Re(A)$ and fluorescence to $F$, as seen in Eq.~\eqref{eq:moleculespectra}.
From the real part of $A$, the imaginary part may be found numerically by Hilbert transform (due to Kramers--Kronig relations).
The response function is then determined from the plasmon eigenfrequency $\omega_c$ and its linewidth $\kappa$ together with the strong coupling constant $g_N$.
Although $g_N$ (and the magnitude of $A$) is unknown, it can be fixed so that it corresponds to a given Rabi splitting.
Lastly, the coupling coefficients $C_{s/p}$ and $\kappa_o^T$ are needed to evaluate the spectra.
However, if we are only interested in the relative magnitudes, it is enough to fix the ratios $\kappa_o^T/C_s$ and $C_p/C_s$.
The latter ratio is given by the polarization angle $\beta$ which can be evaluated with the dielectric functions of the materials at the interface where the plasmon is excited.
The former ratio $\kappa_o^T/C_s$ is difficult to determine directly from the experiments but it can be found by fitting to experimental data.

In Fig.~\ref{fig:comparison} we employ the above method to compare the experimental results for the polarization ratio of TDBC and R6G molecules from Ref.~\onlinecite{baieva2017dynamics} to our model.
Here, the polarization ratio $P^T_p/P^T_s$ of lower (upper) polariton is defined as the ratio of the p- and s-polarized emission peak intensity of lower (upper) polariton.
In our numerical analysis, we approximate the fluorescence data by a mirror image of the absorption data over the zero-phonon frequency $\tilde \omega_m$.
We assume that the fluorescence is effectively independent of the driving so that we can calculate and consider only the elastic emission (see also Appendix~\ref{sec:correspondence}).
Adding the inelastic emission can only diminish the polarization ratio.
Then we calculate the polarization ratio using Eq.~\eqref{eq:scemission} and fit the coupling rate ratio $\kappa_o^T/C_s$  to the experimental data for each branch separately.

For lower polariton peaks we find reasonable agreement with the fitted theoretical curves and the experimental data. 
For upper polariton the correspondence is very limited.
Theoretically, we would assume that the polarization ratio increases for the upper polariton for positive detunings while for the lower polariton the ratio increases for negative detunings.
This is caused by the polaritonic state becoming more plasmonic,
which is seen from the response function being peaked at $\omega_d \approx \omega_c$ in Fig.~\ref{fig:rfs}.
While these trends can be seen in Fig.~\ref{fig:comparison}, except for R6G upper polariton branch, some features differ from the theoretical description.
From experimental point of view the polarization ratio might be affected by any external noise especially in regions of small s-polarized emission.
In our modeling we neglect the possible dependence of the plasmon linewidth and coupling rate on the plasmon eigenfrequency.
Also, since we use the experimental absorption data to determine the spectra the lineshape far from the absorption maximum is also important.

The fitted ratio $\kappa_o^T/C_s$ controls, generally speaking, the magnitude of the polarization ratio.
The effect of Stokes shift seems to be important only at plasmon eigenfrequencies below the fluorescence frequency of the molecules while the external coupling rates control the polarization ratio at higher frequencies.
From the fit to the experimental data we find that the ratio $\kappa_o^T/C_s$ is larger for R6G than for TDBC.
This implies that if the plasmonic emission rate remains the same for TDBC and R6G samples, there is more molecular emission for TDBC than for R6G.

\subsection{Coherent polaritonic response}

If we assume that the plasmon couples strongly to the molecules and drives them coherently, the phase factor $e^{i \vec k \cdot \vec r_j}$ is fixed to a constant.
This leads to the introduction of two different sums over the molecular indices
\begin{equation}
\tilde C_{s/p} = \sum_{j} g_j \eta_{j,s/p}^{T/R}  =
        \begin{cases} 
             0  \\
              g_N\sqrt{\frac{N}{3}\kappa_m^{T/R}}\sin{\beta}, 
        \end{cases}
\end{equation}
where again the upper/lower line is for s/p.
The sum in $\tilde C_s$ vanishes because 
the plasmon polarization vector $\hat u_{pl}$ is orthogonal to the s-polarization vector $\hat u_s$ making the product antisymmetric under reflection through the plane orthogonal to $\hat u_s$.
This is what breaks the symmetry between the polarization directions emitted from the strongly coupled mode.

Then, we find the polarized emission spectra to be
\begin{subequations}
\begin{align}
    S^T_s(\omega;\omega_d) &= \abs{\alpha r(\omega_d)}^2  C_s \qty[\Ione - \abs{\Itwo(\Delta)}^2\delta(\omega - \omega_d)]
	\label{eq:semission:coh}\\
    S^T_p(\omega;\omega_d) &=  \abs{\alpha r(\omega_d)}^2 \kappa_o^T \abs{1 + i \frac{\tilde C_p}{\sqrt{\kappa_o^T}} \Itwo(\Delta)}^2 \delta(\omega - \omega_d) \notag\\ 
    &\!\! + \abs{\alpha r(\omega_d)}^2 C_p \qty[\Ione - \abs{\Itwo(\Delta)}^2\delta(\omega - \omega_d)].
    \label{eq:pemission:coh} 
\end{align}
\end{subequations}
The s-polarized emission vanishes now fully when the vibrations are absent.
This result shows that coherence is crucial to the destructive interference of emitted light \cite{baieva2017dynamics} while the vibrations still provide a mechanism for a partial loss of coherence. 
On the other hand, the p-polarized spectrum contains the interference terms between the plasmon and molecular output fields.
Similar to the Stokes shift case, there are terms with different powers of the number $N$ of molecules.
Considering the Rabi splitting (or $g_N$) to be fixed, there is one term with an extra $N$ factor from ${\tilde C_p}^2$ and $\sqrt{N}$ from $\tilde C_p$.
Increasing $N$ leads to mostly elastic p-polarized emission, as the inelastic terms and s-polarized emission are independent of $N$.
In contrast to the Stokes shift case, the absence of vibrations does not remove the $N$ dependence.
Therefore, the result corresponds to superradiance \cite{dicke1954coherence}.

\begin{figure}
	\includegraphics{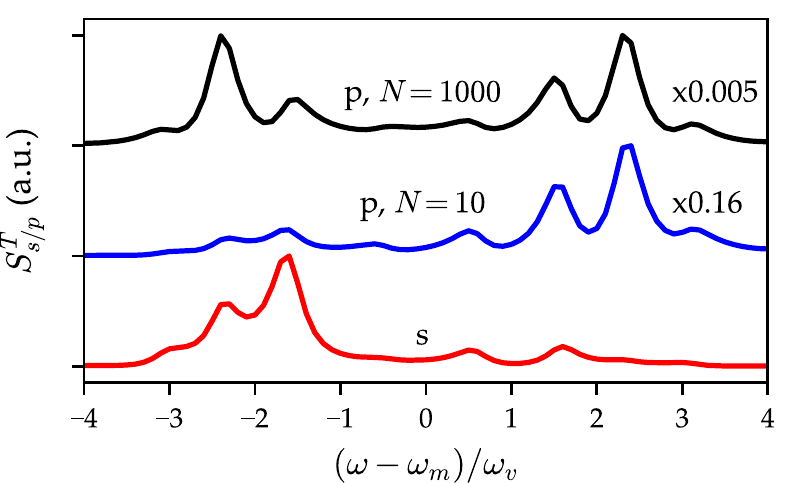}
    \caption{\label{fig:emission:coherent} Elastic emission spectra in the coherent case for two different numbers $N$ of molecules.
    The curves for p-polarized emission (only ones depending on $N$) are offset and scaled.
    Here, $S=1$ and the other parameters are the same as in Fig.~\ref{fig:elemission}.
    }
\end{figure}

Figure \ref{fig:emission:coherent} shows that the coherence and the resulting interference between plasmonic and molecular emission have a qualitative effect on the elastic spectra. 
The difference between s- and p-polarized spectra becomes more evident.
While the s-polarized emission is likely to occur on the lower polariton frequency, for the p-polarized emission the upper polariton frequency may be favored depending on the relative magnitudes of $\kappa_o^T, \tilde C_p$ and $C_p$.

\section{Conclusions}
To summarize, we have constructed a model that allows describing the effect of vibrations on the strongly coupled stationary response of driven coupled light-matter modes. Depending on the case, one can either find the $P(E)$ function describing the absorption and emission of vibrations in a given model system, or relate the measured absorption and fluorescence of uncoupled molecules to $P(E)$. With small modifications, this approach can be extended also to the case of molecule-cavity systems \cite{yu2019strong,canaguier2013thermodynamics,schwartz2013polariton,Chovan2008PRB,Virgili2011PRB}, plasmonic lattices \cite{hakala2018bose} and/or higher-order correlation functions of the emitted light \cite{wang2019turning}. Our quantum Langevin equation approach allows describing the stationary driven system, and hence it complements the often-used computational methods usually concentrating on transient response \cite{delpino2018tensor, Groenhof2019JPCL}.

\begin{acknowledgments}
We acknowledge the support from the Academy of Finland Center of Excellence program (project no. 284594) and project numbers 289947, 290677, and 317118.
\end{acknowledgments}

\appendix

\section{Generalization to many nonidentical and interacting vibrational modes}\label{sec:multiplemodes}
The $P(E)$ theory is straightforward to generalize to multiple vibrational modes when the modes couple linearly to the molecule. 
A general interaction term in the Hamiltonian is then $\lambda_{ijkl} b_{ij}^\dagger b_{kl} + \text{h.c.}$ where $b_{ij}$ corresponds to the $j$th vibrational mode of the $i$th molecule. 
The molecule-vibrational Hamiltonian is then diagonalized by first diagonalizing the vibrational Hamiltonian and then using the polaron transformation.
For simplicity, let us now discuss the case of a single molecule.
After the diagonalization of the vibrational part, we may write the interaction Hamiltonian in terms of the new diagonal vibrational modes $b_j$ as
\begin{equation}
    H_{m+v} =  \sum_{j=1}^M \omega_{v,j} \sqrt{S_j} \sigma^\dagger \sigma \qty( x_j + u_j p_j),
\end{equation}
where $x_j$ and $p_j$ are the position and momentum operator of vibrations.
The term $u_j p_j$ follows from the fact that the molecule couples to the bare vibrational modes.
Because in the Caldeira--Leggett model the position operator $x_j$ couples to the position operator of an environmental (harmonic) mode, the diagonalization is incommensurate with this model unless $u_j=0$.
In the single-excitation limit we may then introduce the operator $\sigma^S = Q \sigma \equiv \prod_j Q_j \sigma$ where $Q_j = e^{\sqrt{S}\qty(b_j^\dagger - b_j)}$.
Introducing many molecules into the situation only adds one external index to each operator.
When there is no coupling to the plasmon, by following the same approximations as in the main text, we find that the dynamics is given by the input-output equation
\begin{equation}
	\dot \sigma^S = - \qty(i \tilde{\omega}_m + \frac{\kappa_m}{2})\sigma^S - \sqrt{\kappa_m^{\rm ext}} Q\sigma_{{\rm in}}
\end{equation}
with $\tilde\omega_m = \omega_m - \sum_j S_j\omega_{v,j}$.
The equation for $b_j$ again decouples from the dynamics of $\sigma^S$ in the single-excitation limit.
Similarly to the case of a single vibrational mode we find the two- and four-point correlators of $Q$ in the calculation of the spectra.
However, since $Q(t) = \prod_j Q_j(t)$, the Fourier transform of $Q$ is always a convolution.
After diagonalization we may treat the modes as independent so this structure shows up as convolutions of $P(E)$'s and $L$'s defined in the earlier sections.
Thus, we have
\begin{widetext}
\begin{subequations}
\begin{align}
	&\expval{Q^\dagger(\omega_1) Q(\omega_2)} =  P_{\rm tot}(\omega_1) \delta(\omega_1 + \omega_2) \\
    &\expval{Q^\dagger(\omega_1) Q(\omega_2)  Q^\dagger(\omega_3) Q(\omega_4)}  = L_{\rm tot}(\omega_1,\omega_2,\omega_4) \delta(\omega_1 +\omega_2 + \omega_3 + \omega_4),
\end{align}
\end{subequations}
where $P_{\rm tot}(E) = [P_1 * P_2 * \dots * P_M](E)$ is a convolution over $M$ different modes and similarly for $L_{\rm tot}$.
For example,
\begin{equation}
    [L_1 * L_2] (\omega_1,\omega_2,\omega_3) = \int \dd{\omega_1'} \dd{\omega_2'} \dd{\omega_3'} L_1(\omega_1 - \omega_1',\omega_2 - \omega_2',\omega_3 - \omega_3')L_2(\omega_1',\omega_2',\omega_3'),
\end{equation}
where $L_1$ and $L_2$ are defined for a single mode as the Fourier transform of Eq.~\eqref{eq:L:start}.
\end{widetext}

\section{Approximation to the polaron equation}\label{sec:approximation}
We discuss the consistency of the approximation that allows us to simplify the input-output equation of $\sigma^S_j$.
The full dynamical equation of $\sigma^S_j$ for the strongly coupled plasmon--molecule system, using the approach of \cite{gardiner1985input}, is given by
\begin{align}
\dot \sigma_j^S &= - i \tilde{\omega}_m \sigma_j^S + i g_j Q_j \sigma_{z,j}c 
    - \frac{\kappa_j}{2} \sigma_j^S + \sqrt{\tilde\kappa_j} Q_j\sigma_{z,j}\sigma_{\rm{in},j}
    \notag\\
	&+ \frac{\gamma_j\sqrt{S}}{2}\sigma_j^S(b_j -  b_j^\dagger) 
    + \sqrt{\gamma_j S} \sigma_j^S(b_{{\rm in},j} - b_{{\rm in},j}^\dagger).
    \label{eq:fullpolaron}
\end{align}
In the main text, we assumed the single-excitation limit in which $\sigma_{z,j}\approx -1$.
In addition, we neglect the thermal fluctuations and set $\sigma_{{\rm in},j} = 0$.
Next we discuss when we can neglect the two last terms that are generated by the coupling of $\sigma^S_j$ to the the vibrational baths.
This approximation effectively uncouples the vibrational dynamics from the dynamics of the polaron operator $\sigma^S_j$.
As this approximation is related to the molecule--vibration system, also the coupling to the plasmon may be neglected, i.e. $g_j = 0$.
For notational brevity, we omit the molecular index $j$.
Let us consider an expansion $\sigma^S =  \sigma^S_0 + \tilde\sigma^S_1$, where $\sigma^S_0$ is the solution of
\begin{equation}
    \dot \sigma^S_0 = - \qty(i \tilde \omega_m + \frac{\kappa_m}{2})\sigma^S_0 - \sqrt{\kappa^{\rm ext}_m}Q \sigma_{\rm in},
    \label{eq:approx1}
\end{equation}
on Eq.~\eqref{eq:fullpolaron} with the above simplifications.
Consequently, the dynamics of $\tilde\sigma^S_1$ is given by
\begin{widetext}
\begin{equation}
    \dot{\tilde\sigma}^S_1 = - \qty(i \tilde \omega_m + \frac{\kappa_m}{2})\tilde\sigma^S_1
    +\frac{\gamma\sqrt{S}}{2}(\sigma^S_0 + \tilde\sigma^S_1)(b -  b^\dagger) 
    + \sqrt{\gamma S} (\sigma^S_0 + \tilde\sigma^S_1) (b_{{\rm in}} - b_{{\rm in}}^\dagger).
    \label{eq:approx2}
\end{equation}
We may now construct the next order of the expansion by setting $\tilde \sigma^S_1 = \sigma^S_1 + \tilde \sigma^S_2$ and fixing $\sigma^S_1$ to be the solution of
\begin{equation}
    \dot \sigma^S_1 = - \qty(i \tilde \omega_m + \frac{\kappa_m}{2})\sigma^S_1 
    +\frac{\gamma\sqrt{S}}{2}\sigma^S_0(b -  b^\dagger) 
    + \sqrt{\gamma S} \sigma^S_0 (b_{{\rm in}} - b_{{\rm in}}^\dagger).
    \label{eq:approx3}
\end{equation}
This equation can be solved with the solution of $\sigma_0^S$.
The dynamics of $\tilde \sigma^S_2$ is then determined by an equation similar to Eq.~\eqref{eq:approx2} where $\tilde \sigma^S_1$ is replaced by $\tilde \sigma^S_2$ and $\sigma^S_0$ by $\sigma^S_1$. 
Continuing this process gives then the expansion of $\sigma^S = \sum_{j=0}^\infty\sigma^S_j$.
However, we focus only on the first order of the expansion.

Consider now that the molecule is driven coherently, $\sigma_{\rm in} = \alpha e^{-i \omega_d t}$ so that the solution of Eq.~\eqref{eq:approx1} is
\begin{equation}
    \sigma_0^S(\omega) = \alpha \sqrt{\kappa^{\rm ext}_m} \chi(\omega - \tilde\omega_m)Q(\omega - \omega_d).
\end{equation}
Consequently, we obtain from Eq.~\eqref{eq:approx3}
\begin{equation}
    \sigma^S_1(\omega) = \sqrt{S}\chi(\omega - \tilde\omega_m) 
    \qty(i\frac{\gamma}{2} \qty[\sigma_0^S * p](\omega) + \sqrt{\gamma}\qty[\sigma_0^S * \qty(b_{{\rm in}} - b_{{\rm in}}^\dagger)](\omega)),
\end{equation}
where $*$ denotes a convolution in the Fourier space.
\end{widetext}

We are now interested in the consistency of the expansion but it is not straightforward to see the effect of the convolution and the underlying dynamics of the vibrations.
For this reason, we compare the mean values of $\sigma^S_0$ and $\sigma^S_1$.
When the input operators of the vibrations represent thermal noise, the $b_{\rm in}$ terms do not contribute to the average.
The expectation value of $\sigma^S_0$ can be expressed as
\begin{equation}
    \expval{\sigma^S_0(\omega)} =  \alpha \sqrt{\kappa^{\rm ext}_m} \chi(\omega - \tilde\omega_m)\expval{Q(0)}\delta(\omega - \omega_d).
\end{equation}
For a thermal ensemble $\expval{Q(0)} = \exp(-S(n_{\rm th} + \frac{1}{2}))$.
In the calculation of the average of $\sigma^S_1$ we need the generalized Wick theorem to write
\begin{equation}
    S\expval{p^n(t)p(0)} = n J(t) \expval{p^{n-1}(t)}
\end{equation}
where $J(t)$ is the function defined in Eq.~\eqref{eq:Joft} and we define its Fourier transform by $J(t) = \int \dd{\omega} e^{-i \omega t}J(\omega)$.
We obtain
\begin{widetext}
\begin{align}
    \expval{\sigma^S_1(\omega)} &= \frac{\gamma}{2} \alpha \sqrt{\kappa^{\rm ext}_m} \chi(\omega - \tilde\omega_m) \qty[\int \dd{\omega'} \chi(\omega' + \omega_d - \tilde\omega_m)J(\omega')] \expval{Q(0)} \delta(\omega - \omega_d) \notag \\
    &= \frac{\gamma}{2} \qty[\int \dd{\omega'} \chi(\omega' + \omega_d - \tilde\omega_m)J(\omega')] \expval{\sigma^S_0(\omega)} \equiv \mathcal{C}\expval{\sigma^S_0(\omega)}.
\end{align}
Now we have a necessary condition for the consistency of the simplification: The parameter $\mathcal{C}$ should be small compared to unity for the expansion to be sensible.
It can be estimated by using the same approximation in Eq.~\eqref{eq:twopeakLorentz} as in the $\gamma = 0$ calculation.
Then (denoting $\Delta = \omega_d - \tilde \omega_m$)
\begin{equation}
    \mathcal{C} = \frac{\gamma S}{2} \qty[(n_{\rm th} + 1)\chi(\omega_v + \Delta) + n_{\rm th}\chi(-\omega_v + \Delta)]
    \overset{\Delta = -\omega_v}{\approx} \frac{\gamma S}{\tilde\kappa + \gamma S} \frac{n_{\rm th}+1}{2}.
\end{equation}
\end{widetext}
In the last approximation we have written the renormalized linewidth $\kappa_m$ in terms of the bare linewidth of the molecule $\tilde\kappa$ and neglected the smaller term $\chi(-2\omega_v)$ for clarity.
Now, it is clear that the consistency of the approximation is related to the temperature and the linewidths.
This condition is always fulfilled when $n_{th} < 1$ or alternatively $\frac{\omega_v}{k_B T} > \ln(2) \approx 0.69.$
It should be remembered that this is only a crude estimate and larger values of $\gamma$ can diminish the value of $\mathcal{C}$.

\section{Correspondence to experiments}\label{sec:correspondence}
In the experimental Kretschmann setup a prism is used together with white light to excite the plasmons.
The angle $\theta$ of incoming light with respect to the normal of the interface then determines the plasmon eigenfrequency $\omega_c$.
In our model, which is based on a coherent single frequency driving at $\omega_d$, we can introduce a distribution $\rho(\omega_d)$ for light intensity.
Then we can relate our theoretical model to the observed spectra by an integral relation
\begin{equation}
    S^{T/R}_{s/p,\mathrm{obs}}(\omega) = \int \dd{\omega_d} \rho(\omega_d) S^{T/R}_{s/p}(\omega;\omega_d).
\end{equation}
The distribution $\rho(\omega_d)$ can include features of the driving light as well as the prism that couples the light to the interface.
We can generally divide the theoretical spectrum into elastic and inelastic part by $S^{T/R}_{s/p}(\omega;\omega_d) = S_{\mathrm{el}}\delta(\omega - \omega_d) + S_{\mathrm{inel}}(\omega;\omega_d)$.
Now, if we assume that the distribution of light is uniform and its bandwidth large compared to the plasmonic linewidth $\kappa$ we have
\begin{equation}
    S^{T/R}_{s/p,\mathrm{obs}}(\omega) = S_{\mathrm{el}}(\omega) + \int\dd{\omega_d}S_{\mathrm{inel}}(\omega;\omega_d).
\end{equation}
The inelastic contribution is due to vibrations and molecular fluorescence (i.e. the function $F(\omega \neq \omega_d)$).
In the fit for Fig.~\ref{fig:comparison}, we disregard the contribution from $S_\mathrm{inel}$, because its main body is clearly separated from the elastic emission coming around the polariton frequencies.


\begin{thebibliography}{54}
\bibitem [{met()}]{methodnote}%
  \BibitemOpen
  \href@noop {} {}\bibinfo {note} {Possible alternative approaches include the
  Lindblad master equation approach for the density matrix, or the stochastic
  quantum jump approach. The previous would have to be written in the polaron
  basis to allow for a proper treatment of the effect of vibrations. The latter
  is most suitable for computational approaches.}\BibitemShut {Stop}%
\end{thebibliography}
\end{document}